\documentclass{article}

\usepackage{arxiv}

\usepackage[utf8]{inputenc} 
\usepackage[T1]{fontenc}    
\usepackage{hyperref}       
\usepackage{url}            
\usepackage{booktabs}       
\usepackage{amsfonts}       
\usepackage{nicefrac}       
\usepackage{microtype}      
\usepackage{lipsum}		
\usepackage{graphicx}
\usepackage[square,numbers,sort]{natbib}
\usepackage{doi}

\usepackage{graphicx}
\usepackage{algorithmic}
\usepackage{amssymb}
\usepackage{amsmath}
\usepackage{subfigure}
\usepackage{csquotes}

\title{Generating Local Maps of Science using Deep Bibliographic Coupling.}

\author{
	\href{https://orcid.org/0000-0002-9072-1535}{\includegraphics[scale=0.06]{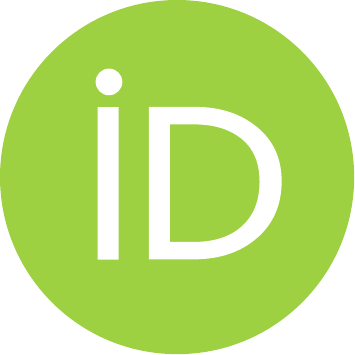}\hspace{1mm}Ga\"elle Candel} \\
	Wordline TSS Labs, Paris \\
	\texttt{firstname.lastname@worldline.com} \\
	\& \\
	D\'epartement d'informatique de l'ENS \\
	ENS, CNRS, PSL University, Paris \\
	\texttt{firstname.lastname@ens.fr}\\
	\And
  David Naccache \\
	D\'epartement d'informatique de l'ENS \\
	ENS, CNRS, PSL University, Paris \\
	\texttt{firstname.lastname@ens.fr}\\
}

\date{}

\renewcommand{\headeright}{}
\renewcommand{\undertitle}{}
\renewcommand{\shorttitle}{Generating Local Maps of Science}

\hypersetup{
pdftitle={Generating Local Maps of Science using Deep Bibliographic Coupling.},
pdfsubject={cs.AI, cs.DL},
pdfauthor={Ga\"elle Candel, David Naccache},
pdfkeywords={Bibliographic Coupling, Citation graph, Community Detection, Research Front, Visualization},
}

\begin{document}
\maketitle

\begin{abstract}
	Bibliographic and co-citation coupling are two analytical methods widely used to measure the degree of similarity between scientific papers.
	These approaches are intuitive, easy to put into practice, and computationally cheap.
	Moreover, they have been used to generate a map of science, allowing visualizing research field interactions.
	Nonetheless, these methods do not work unless two papers share a standard reference, limiting the two  papers’ usability with no direct connection.
	In this work, we propose to extend bibliographic coupling to the deep neighborhood, by using graph diffusion methods.
	This method allows defining similarity between any two papers, making it possible to generate a local map of science, highlighting field organization.
\end{abstract}


\section{Background}

\subsection{History and Usage of Coupling Metrics}

Historiographic mapping has been introduced by Garfield \cite{Garfield2004HistoriographicMO}, retracing the  connections between the different discoveries around DNA.
This mapping of historical events was made possible thanks to Bibliographic Coupling (BC), introduced by \cite{Kessler1963BibliographicCB}, and Co-citation Coupling (CC), introduced by \cite{Small1973CocitationIT}.
Both methods use the same approach: two documents are similar if they share at least one cited or citing a paper in common.
The similarity score is a function of the overlap between the sets of references or referees.
These coupling methods have been used extensively for many purposes: to create a map of science, to study relationships between universities \cite{Grauwin2011MappingSI}, to study relationships between fields \cite{leydesdorff2013scientometrics}, to analyze co-authorship networks \cite{Ding2011ScientificCA}, or to retrace phylogeny of science \cite{Boyack2005Backbone}.

\subsection{Formal Representation of Citation Graphs}

\paragraph*{Directed Acyclic Graph:}

A citation graph can be represented as a directed acyclic graph (DAG) $G = (V, E)$, with $E \subseteq V \times V$, where vertices $V$ represent published papers and edges $E$ the citation relationships, produced from the most recent paper to the oldest one.
The formed graph corresponds to a DAG, as it is not possible to find a path from a node to itself, as the edges are directed toward the past.

\paragraph*{References and Citations}

The reference set is denoted $R(a) = \{b | (a, b) \in E\}$ for a document $a$ and its citation set is denoted similarly as $C(a) = \{b | (b, a) \in E\}$.
This set corresponds to all reachable items at a distance $1$ from the currently studied paper $a$ and can be denoted $R^1(a) = R(a)$.
The reference set can be recursively extended to indirectly referenced items reachable within $k$ steps from $a$ as  $R^k(a) =  \bigcup_{b \in R^1(a)} R^{k-1}(b) \cup R^1(a)$.
The citation set can be extended the same way as $C^k(a) =  \bigcup_{b \in C^1(a)} C^{k-1}(b) \cup C^1(a)$.

\paragraph*{Couplings}

Two papers are coupled if they share some of their neighbors.
For BC, the set of papers coupled with $a$ is denoted $BC(a) = \bigcup_{b \in R(a)} C(b)$, while $CC(a) = \bigcup_{b \in C(a)} R(a)$ for co-citation.
The coupling can be extended by including indirect neighbors up to distance $k$ as
$BC^k(a) = \bigcup_{b \in R^k(a)} C^k(b)$ and $CC^k(a) = \bigcup_{b \in C^k(a)} R^k(a)$.

\paragraph*{Scoring Relationships}

The main similarity measures used by BC and CC are the \textit{Cosine} and \textit{Jaccard} similarity measures.
Both measure the relative overlap between two sets $A$ and $B$ but distinguish by the normalization.
The cosine similarity is defined as:

\begin{equation}
  \label{eq:cosine}
  S_C(A, B) = \frac{|A \cap B|}{\sqrt{|A|.|B|}}
\end{equation}

and the Jaccard similarity as:

\begin{equation}
  \label{eq:jaccard}
  S_J(A, B) = \frac{|A \cap B|}{|A \cup B|}
\end{equation}

Their behavior is relatively the same.
They are both equal to $1$ for perfect overlap ($A = B$) and equal to $0$ for disjoint set ($A\cap B = \emptyset$).
They only differ in the numerical value for the intermediate cases.
For two BC coupled items $a$ and $b$ considering their indirect references up to distance $k$,
the similarity score using the cosine similarity is measured as $S^k_{BC\text{-}C}(a, b) = S_C\left(R^k(a), R^k(b)\right)$.

\subsection{Bibliographic Coupling vs Co-Citation Coupling}

BC and CC are relatively similar from a graph perspective.
BC looks at the references (outgoing edges)  while CC looks at the citations (incoming edges).
The approaches are precisely the same with the difference that the direction of the edges is reversed for CC.
Several works have been done to compare the two approaches \cite{Boyack2010CocitationAB,Klavans2017WhichTO}.
In the next paragraphs, we will present some of the differences.

\subsubsection{Article Decomposition:}

An article starts with a bibliographic overview, with an \textit{introduction} setting the scope of the field and introducing the problem.
Then, \textit{background}, \textit{related works} or \textit{state-of-the-art} sections referencing related methods solving the current problem or a similar one, spot missing points or possible transposition to the current problem that the paper will address.
The following part corresponds to the authors’ contributions, exposing a novel method supported by an experimental protocol, results, and analysis.
The first part presents authors' arguments supported by \textit{bibliographic} evidence, while the second part presents authors' ideas supported by \textit{numerical} evidence.

Most of the time, the article format is imposed on the authors.
Therefore, there is a trade-off between the two parts as space is limited.
An extensive bibliographic section supports the article with strong argumentation, to the detriment of the main work.
On the other hand, a problem introduction that is too short limits the understanding of the working context, the problem and the potential impact.
In general, the second part is relatively more extensive than the first, expect for \textit{"reviews of"} where the aim is to compare the existing literature widely.

\subsubsection{Evolution over Time:}

A citation graph is a dynamic structure where new nodes are added over time.
The reference count of an article will not change over time (unless the database is consolidated with old documents).
However, the number of citations received by an article starts at $0$ and will increase over time.
This difference between the evolution of $|R(a)|$ and $|C(a)|$ impacts the usage of BC and CC.

An article is bibliographically coupled with other articles at the publication date, unless it has no reference or the quoted papers have never been cited before.
Over time, the number of coupled articles will increase as other nodes will refer to its references.
The database may not cover very old articles or documents with different types (like patents, blogs, and others).
For the nodes referring to these unregistered items, their effective reference set might be incomplete.

In contrast, an article is not co-citation coupled with any other article at the publication date, as no paper could mention this work yet.
The citation count will increase over time, increasing the number of coupled items.
The coupling strength between two items will vary over time as their respective citation set will evolve.
The strength increases if they are mentioned together or decreases if cited separately.

BC is not adapted for very old papers, while CC is not for very recent ones.
For not-too-old and not-too-recent papers, they can both be used.
BC has the advantage to be usable for freshly published work and even for unpublished works.
Therefore, BC is useful to study research fronts or analyze a work under review by looking at its references.

\subsubsection{Degree asymmetry:}

The nodes' degree distribution differentiates the evolution of $|BC(a)|$ from $|CC(a)|$.
The number of references per article has a low variance, while the variance is much higher for citations.
Citations are distributed according to a power-law, where a few works receive many citations and the large majority almost none \cite{Price1976AGT,Redner1998HowPI}.
One theory that could explain this phenomena is the \textit{Matthew effect}, which could be summarized as "the rich get richer and the poor get poorer" \cite{Merton1968TheME}.
The works that already have many citations will get more than works that have already fewer citations.

The \textit{Matthew effect} tends to model the long-term evolution of $|C(a)|$ as a function of citations' actual number.
At the publishing date, all papers are equivalent as they have no citation yet.
Different factors impact the initial citation rates of newly published papers.
There are \textit{notoriety} factors, such as the publication context (prestigious journals and conferences), the author's affiliation and the author's reputation \cite{Merton1968TheME},
and \textit{visibility} factors, where the authors could present their work in university, publish a vulgarized version for broader dissemination \cite{Allen2013SocialMR}, or write new papers exploiting their work, and open-access \cite{Gargouri2010SelfSelectedOM}.

This phenomenon impacts more co-citation analysis than bibliographic coupling
as the number of coupled items with BC depends on the references' fame.
In contrast, it depends on the paper's citation count for CC coupling which represents a bottleneck.

\subsubsection{Conceptual Differences between BC and CC}

While the approaches are relatively similar, they differ in their interpretation.

\paragraph{Citation Control:}
Referencing a paper is an \textit{active} action.
The authors take their time to look at the literature, review several articles to select some of them to support their argumentation.
The number of references is indirectly limited by the paper format, where an extensive reference list would limit the space available to present the main paper's contribution.
Therefore, references must be selected carefully to support the argument concisely.

To be cited is a \textit{passive} process.
As a first step, authors can actively contribute to disseminating of their work, by presenting it to their peers, during workshops and other seminars.
This would increase the visibility of their work, increasing the likelihood of receiving citations.
However, the quantification of all formal and informal interactions the authors had is intractable with the available tools.

Out of the presentation of their work to a targeted audience, the authors have no direct control over who will cite their work.
It could be people from the same field, proposing an extension of the current work with new ideas, or people from another field transposing the proposed idea to their problem.
Thus, the citation process can be seen as \textit{crowdsourcing}, as an article will be referenced based on its usefulness and not necessarily on the context to which it refers.
Therefore, co-citation analysis will give larger scores to items with a similar impact independently of their initial field.

\paragraph{Field Asymmetry:}
In theory, BC and CC are equivalent for a paper far from the graph boundaries.
The number of coupled items can be considered  equivalent for both BC and CC if there is a sufficient number of citations and references.
It is unlikely that $BC(a)$ and $CC(a)$ broadly overlap because of the graph sparsity and the large number of possible citation choices.
Nonetheless, it is a reasonable assumption that they gather items belonging to the same research domain.
This might be true most of the time, but a few exceptions exist.
One example illustrating the possible asymmetry is \textit{“TensorFlow: a system for large-scale machine learning”} \cite{AbaBar16Tensorflow}.
This paper discusses \textit{distributed computation systems} in its core text and references, which could be used for \textit{Deep Learning}.
However, most citations come from Deep Learning papers rather than from distributed systems.
This asymmetry could be explained by the largest impact of the technology on \textit{Deep Learning} than on the \textit{distributed system}.

The use of co-citation analysis is more adapted to study the impact of papers, for historiographic purposes.
However, this method is not adapted for non-top papers or recent papers as the citation count impacts the coupling highly.
In contrast, the bibliographic coupling can deal with  high to low quality papers as long as they mention relevant literature, allowing to study scientific communities in their globality and active research fronts.
Therefore, we will focus on the BC analysis for the rest of the paper.

\section{References Selection and its Impact on Coupling Strength}

In this section, we will discuss  the impact of a reference over the coupling between two papers.
First, we will underline the problems using first-order references $R^1(a)$ and high-order references $R^k(a)$.
Then, we propose an intermediate solution that considers references' contribution discrediting irrelevant papers to the profit of relevant ones.

\subsection{Cited and Un-cited Works}

Authors have to select a limited set of meaningful references to support their work.
We detail in the following some of the reasons governing the choice of a reference rather than another.

\paragraph*{Awareness:}

The first source of omission is the absence of awareness about previous related works.
The recentness of a related work limits its propagation through the scientific community.
Another reason concerns the vocabulary disparity between two fields.
The vocabulary between two field is not necessarily the same; therefore an idea can be described differently.
This prevents textual search engines from gathering the two related ideas.
The same limitation occurs when papers are published in different languages.

\paragraph*{Confidence:}

The second source of omission is the lack of confidence.
ARelevant work may exist, but the authors may not have a sufficient background to understand all the details.
The author can also be doubtful about the methods or the obtained results and may prefer not to mention this work.

\paragraph*{Notoriety:}

A factor impacting citation selection is \textit{notoriety}, which contributes to the \textit{Matthew effect}.
Out of the fact that top-research or top-journal works are more visible, researchers give them more credit \cite{Tahamtan2019WhatDC}.
The best conferences and journals filter papers by expert review in the field only to select best papers.
This stamping allows non-expert people to trust the quality of the work by delegating the analysis to a third party.

\paragraph*{Knowledge Establishment}

Major works are cited religiously, like Kerckhoffs \cite{lacryptomilitaire} for cryptography, or Principal Component Analysis \cite{PCA} for Mathematics.
However, over time, the knowledge is admitted by the field, and there is no need to refer to it, as they tend to become axioms known by everybody.
Therefore, the number of citations they would receive per year will vanish over time.
The same vanishing process also happens to the theories that are invalidated or outdated by new research.
Because those papers are too commonly cited, they are not discriminative enough and increase the computational cost for identifying strongly coupled articles.
Therefore, those papers add more processing difficulties with limited benefit.

\paragraph*{Independent Discoveries}

\textit{Multiple independent discoveries} \cite{Garfield1980IndependentDiscovery} are similar discoveries that occur on a limited time interval, in different places, without interaction between the two (groups of) researchers.
The theory behind these events is that these discoveries are the logical continuum of previous evolution.
The findings happened because the current technological advancement made them possible to emerge naturally.
As discovery takes time to diffuse and to be accepted, the second group of researchers might not be aware of a previous existence.
When two similar works are published, papers following them need to select one or another or both.
Whatever the next papers’ referencing choice, the two candidates would share the same basis or axioms.
At some point, the literature supporting the two works will converge to the same set of concepts and references.
Therefore, choosing one or the other does not modify the supporting evidence.

\subsection{Issues with First-Order References}

Bibliographic coupling and co-citation analysis are straightforward conceptual and practical.
However, these methods suffer from several drawbacks, which limits their usability.
These two approaches explore the first-order references and citations, representing $10 \sim 30$ related works depending on the field, epoch and database.

\paragraph{Link Strength:}
The main issue concerns the \textit{coupling’s strength}.
In the case of bibliographic coupling, coupled items have at least one reference in common.
Nonetheless, the probability of two coupled papers sharing more than one reference is extremely low because of the sparsity.
Thus, coupled strength is more impacted by the total number of references rather than by the number of shared references.
Additionally, the similarity between two items sharing $2$ or $3$ references over $20$, i.e. $10 \sim 15 \%$ (which is a common value in a scientific citation graph),  does not have a great significance.

\paragraph{Sets Overlap:}
BC and CC cannot be used to compare randomly selected items as the probability that their reference sets overlap is almost zero due to the graph sparsity.
Therefore, it is not possible to identify fields proximity directly with BC.

Different approaches have been put into practice to re-mediate these issues.
Some prune papers that are not cited enough \cite{Jarneving2007BibliographicCA,Leydesdorff2009AGM}.
These approaches drastically reduce the number of papers, especially recently published papers that did not receive attention yet.
Therefore, they are not suitable for the technical watch to study the research fronts \cite{Fujita2012DetectingRF,Huang2013DetectingRF}.
Other aggregate papers within a journal \cite{Colavizza2018TheCT} or an institution \cite{Huang2004ConstructingAP,Yan2012ScholarlyNS} level to increase the number of references per entity.
The aggregation also decreases of the number of items to analyze, reducing the overall complexity while getting a coarser view.

\subsection{References' Relevance}

A solution to measure papers similarity without merging them in a larger group is to move from first-order neighbors to higher-order neighbors, which must be done carefully.

\paragraph{Controlling Sets' Size}
Assuming a paper makes on average $\mathbb{E}\left[|R(a)|\right] = d$ references, the reference set size when exploring up to distance $t$ can be estimated as $\mathbb{E}\left[|R^t(a)|\right] = d^t$, without counting possible overlap.
The relationship can be rewritten differently as $\mathbb{E}[|R^t(a)|] = d \times \mathbb{E}[|R^{t-1}(a)|]$, i.e. the exploration one step ahead contribute to a multiplication by $d$ of what was here previously.
This would lead to a set with mostly old papers, which many other papers  would  share.
Therefore, extending to infinity the exploration distance $t$ will inverse the problem of set overlap to a too large overlap, preventing the identification of meaningful pairs.
Additionally, a paper cited as an example coming from another field comes with all its literature independently of its relevance.
This mass of irrelevant documents can counterbalance the papers coming from the main field.
The last issue is the computational cost, which grows with the set size.
To avoid the different issues, the exploration distance $t$ can be selected carefully to limit the set size.
However, with a change from $d^t$ to $d^t \times d$, the error can be large making the correct selection of $t$ difficult.

\paragraph{The Aim of Weighting:}
Items in the reference set can be weighted to limit the contribution of irrelevant papers rather than trying to stop graph reference exploration to the correct value $t$.
This weighting must reflect papers' relevance relatively to the initial paper.
Relevance can be defined as a function of age where the oldest the document is the less relevant it is.
Another way to estimate relevance is to exploit items' reachability from the initial paper.
Among the indirectly cited articles are axiomatic papers to which many references are made.
Even if the initial paper does not reference them, these papers contributed more significantly to the development of the field of knowledge.
In contrast, items that are referenced only once are unlikely to support the paper's ideas strongly.
Thus, the weighting would highlight the axioms while naturally ghosting papers with low relevance to the current subject.

\paragraph{Random Walk Weighting Model:}
The \textit{Random Walk with Restarts} (RWR) is used to weight node contribution for this purpose.
This model explores all the different possible paths from papers to referenced papers, putting a larger weight on papers that could be accessed through multiple paths.
Therefore, more credit is given to references that are multiple times referred within the reference set, representing a literature consensus.
On the other hand, this weighting model discredits papers that are not substantial, like the papers used to illustrate examples but do not contribute to the argumentation, or belong to another field uncommonly cited.
Of course, a different weighting model can be proposed as an alternative (considering papers' popularity for instance).
Still, the Random Walk is one of the simplest ways to discern \textit{interesting} from \textit{irrelevant} articles \cite{Page1999ThePC}.

\section{Measuring Coupling at a Deeper Level}

This section presents the different equations allowing us to measure the similarity between two items using the Random Walk with Restarts.

\subsection{Weighted Similarity Measures}

The cosine similarity can be easily extended to weighted elements.
Assuming that $\mathbf{W} = \{W_c\}_{c \in A}$ are weights associated with the elements in $A$ and $\mathbf{W}'$ to the elements in $B$,
the cosine similarity can be extended to the weighted domain:

\begin{equation}
  \label{eq:cosine_weight}
  S_{w}(A, B, \mathbf{W}, \mathbf{W}') = \frac{\sum_{c \in A \cap B} (W_c . W_c')}{\sqrt{\sum_{c \in A} (W_c)^2  \sum_{c \in B} (W_c')^2}}
\end{equation}

This measure evaluates how well scores attributed in $A$ are similar to scores attributed in $B$.
These similarity measures are equal to $1$ if the sets and associated weights perfectly match and $0$ if $A \cap B = \emptyset$.
This last measure (Eq. \eqref{eq:cosine_weight}) will be used to measure the similarity between two papers' backgrounds.

\subsection{Random Walk with Restarts}

For a paper $a$ and a paper $b \in R^t(a)$, the RWR measures the reachability of $b$ starting from $a$ and is denoted $W_b^a$ .
The algorithm works by exploring the nodes in $R^t(a)$ starting from $a$ by jumping from one node to another using existing outgoing edges.
Explored edges are selected with equiprobability.
The exploration "restarts" in $a$ when a path is long enough (here $t$).
The value can be computed recursively starting with $W_a^a = 1$ and $0$ otherwise by:

\begin{equation}\label{eq:cg_rw_recursion}
  W_b^a = \sum_{c \in C(b)} \frac{W_c^a}{|R(c)|}
\end{equation}

Then two papers $a$ and $b$ with respective weights $\mathbf{W}^a = \{W_c^a\}_{c \in R^t(a)}$ and $\mathbf{W}^b$ can be compared by measuring their weighted overlap  defined as:

\begin{equation}\label{eq:cg_ww}
  S^t_{w\text{BC}}(a, b) = S_{w}(R^t(a), R^t(b), \mathbf{W}^a, \mathbf{W}^b)
\end{equation}

This coupling is non-zero if $a$ and $b$ share some of their background elements.
The strength considers the set overlap and the attributed weights, which can notably change the coupling strength.

\section{Experimental Setup}

\subsection{Dataset Description}

We used for the experiments the DBLP dataset \cite{DBLP} version 12, published in April 2020.
It is a research citation graph,  composed of 4,894,081 scientific papers with 45,564,149 referencing links.
The papers from this database belong to the computer science domain,  a small part of the research landscape. An initial processing is done to keep the largest connected component and remove cycles.
Cycles exist due to some papers'  updates, allowing them to quote papers published after them.
For each paper, the database provides  information such as title, summary, references, authors, and affiliation, plus around ten descriptive keywords about the domain, field, or methods used.

\subsection{Node Sampling}

We propose to construct a \textit{map of science} exploiting the items' similarity.
The computation of a similarity matrix with $n$ elements induces a computational cost of $\mathcal{O}(n^2)$ limiting the usability over very large samples.
Random sampling allows to reduce the sampling size easily, and would allow building a \textit{global} map of science.
Alternatively, targeted sampling exploiting the metadata available in the dataset allows the construction of a \textit{local} map of science.
The keywords associated with papers in the database could be assimilated to a field of study.
Therefore, we select one keyword of interest occurring in a sufficient number of papers.
All papers associated with the keyword are gathered and then sampled at random to obtain the desired sample size.

\subsection{Data Visualization}

To create \textit{a local map of science}, we used the $t$-SNE embedding algorithm \cite{vanDerMaaten2008}.
This embedding algorithm tries to preserve the neighborhood, i.e., similar items are kept close in the embedding, while distant neighbors are kept away.
This algorithm groups similar items, without trying to preserve distances between input and output space.
A distance matrix $D$ is given to the $t$-SNE algorithm, which transforms it into a 2-dimensional representation $Y$.

The similarity obtained for bibliographic coupling $S_{BC}$ and $S_{wBC}$ have the opposite behavior of a distance, as $S=1 \iff D=0$ and $S=0 \iff \lim_{D \rightarrow \infty} D$.
It is commonly accepted to transform $S$ into $D$ by $D = 1 - S$.
However, the range of possible distance values is restricted to $[0, 1]$, limiting expressiveness.
Instead, we propose to transform $S$ into $D$ by inversion:
\begin{equation}
  D = \frac{1}{\epsilon + S} - \frac{1}{\epsilon + 1}
\end{equation}
where $\epsilon$ is a small constant, set to $\epsilon=10^{-5}$, which avoids the division by zero cases.
$\epsilon$  could be considered as the null hypothesis, “what would be the similarity if neighborhoods overlap at random”.
The pipeline can be summarized as:
$$
V \xrightarrow{\text{sampling}} V' \xrightarrow{\text{similarity}} S \xrightarrow{\text{inversion}} D \xrightarrow{t\text{-SNE}} Y
$$

\subsection{Keywords Extraction}

The evaluation of embedding $Y$ is non-trivial as visual appreciation is a subjective evaluation.
As the proposed methodology is entirely unsupervised and not rely on optimizing a specific criterion, there is no objective value to monitor.
Despite this lack of objectivity, we propose to study clusters.
The goal is to check if clusters of papers correspond to thematic clusters, described with a particular set of keywords.

\paragraph*{Extraction based on Relative Frequency:}
\label{section:subkeywords}

A possibility to study keywords is to select the most frequent.
The keyword used for sampling will be first followed by other keywords with large predominance.
Among those frequent keywords, some are frequent in a general context, and some are frequent only in a specific context.
We propose to distinguish between generally frequent keywords from keywords specifically more frequent in the selected domain.
To this purpose, we look at the ratio between local and global frequencies, estimated over $V'$ and $V$.
The ratio is defined as:

\begin{equation} \label{eq:occurrence_ratio}
  R(kw, V, V') = \frac{r(kw, V')}{r(kw, V)}
\end{equation}
where  $r(kw, V)$ represent the frequency of $kw$ within $V$.
A value close to $1$ indicates that the keyword is no more frequent in $V'$ than in $V$, while a larger value indicates that the keyword is specific to this context.

\paragraph*{Extraction based on Mean-Shift:}

The keyword selection process described in \ref{section:subkeywords} is only possible if the selected sample is relatively singular compared to the whole database.
Otherwise, no keyword would be more relevant than another one.
Rather than looking at the predominance of a keyword within the subset, its distribution in the embedding can be analyzed to see if the keyword occurs in localized areas or not to evaluate its specificity.

To this aim, the resulting embedding is clustered using the \textit{Mean-Shift} algorithm \cite{Mean_shift}, configured with a Gaussian kernel with bandwidth $\sigma$.
Then, the embedded items $Y$ are partitioned into $k$ non-overlapping hard clusters $\mathcal{C} = \{C_i\}_{i=1:k}$, where $k$ is automatically found.

Salient keywords are identified for each cluster using TF-IDF.
The frequency of a keyword is obtained by measuring the proportion between its number of occurrences against all other keywords within the documents belonging to the cluster.

\section{Results}

\subsection{$t$-SNE Embedding: Fields Auto-Organization}

A 2D paper visualization is obtained by transforming a similarity matrix using the $t$-SNE embedding algorithm. For our experiments,
we selected $4000$ papers associated with the \textit{Payment} keyword.
Similar experiments have been run using other keywords, and led to similar results and combination of keywords.

\begin{figure}[h]
    \centering
    \subfigure[Bibliographic Coupling (BC)]{\includegraphics[width=0.49\textwidth]{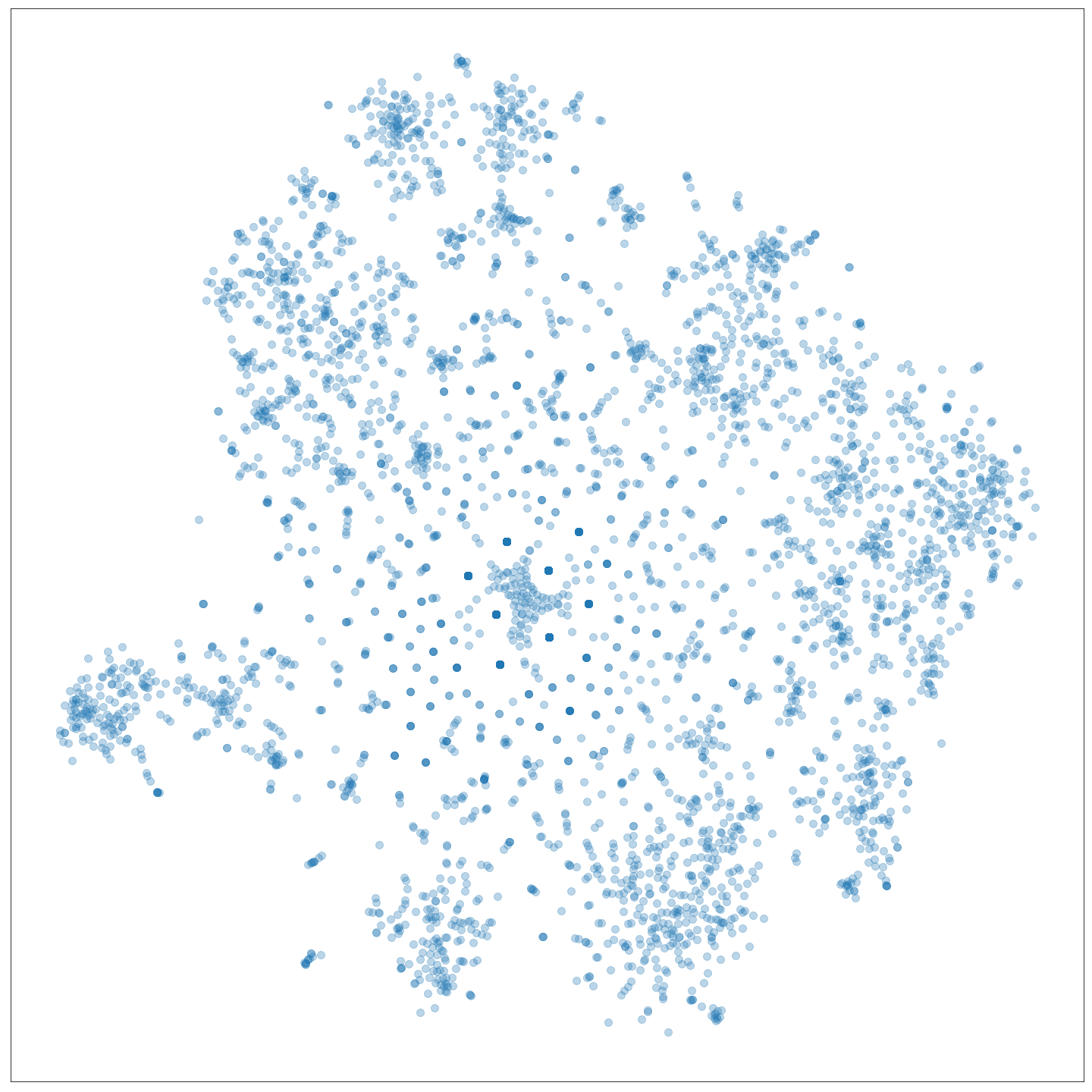}}
    \subfigure[Weighted Bibliographic Coupling (WBC)]{\includegraphics[width=0.49\textwidth]{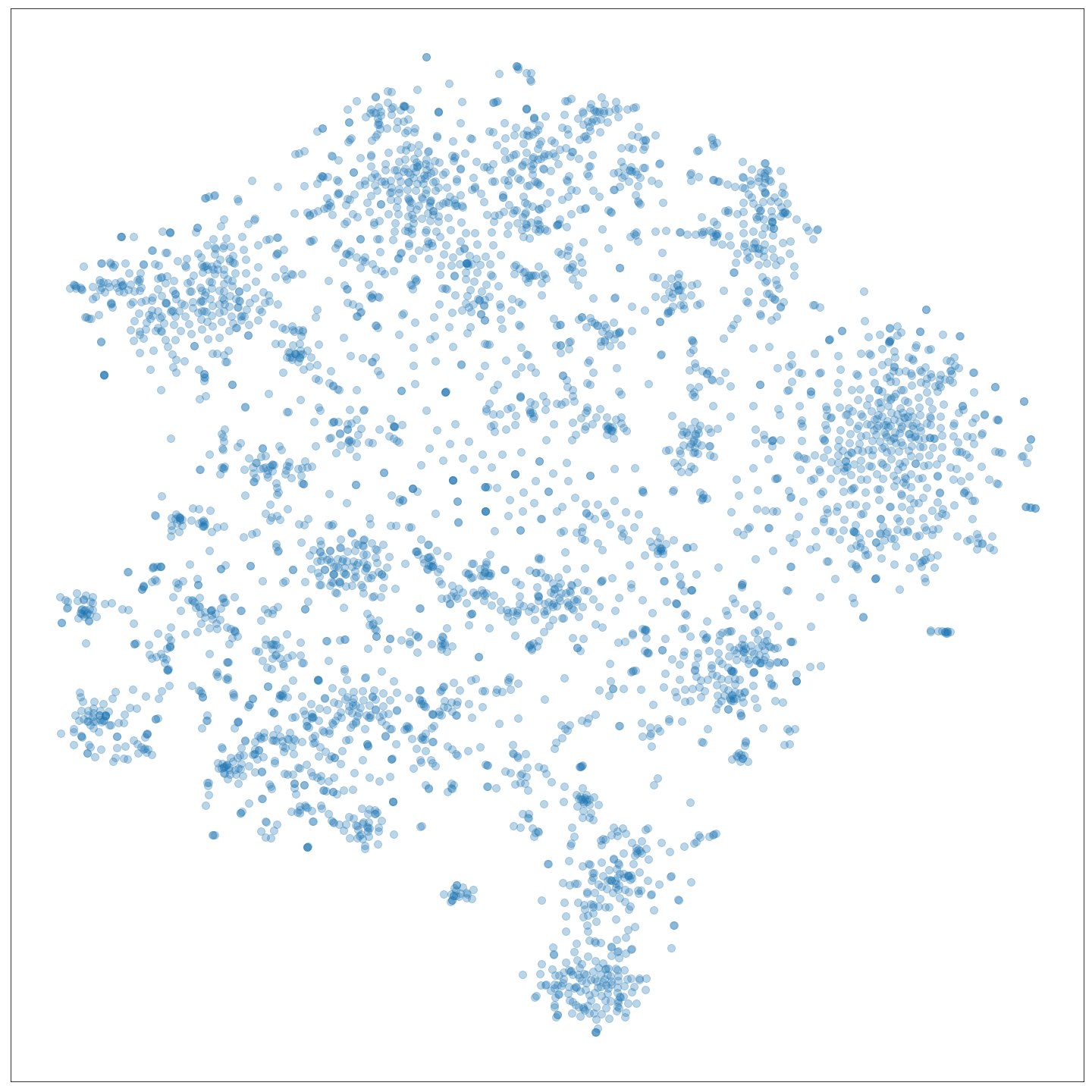}}

    \caption{Local Map of Science for “Payment” papers.
    }
    \label{fig:images_2}
\end{figure}

Initially, embedding results seem quite similar if we don't consider clusters' location.
On both figures, clusters are located on the embedding boundary, while the central areas are less dense with points spaced from each other.
Nonetheless, there are main notable differences.

The main one concerns the number of unclustered items, which is larger in the BC figure than in the WBC one.
These items are easily identifiable as they are located in the middle and are more distant from their nearest neighbors than items in peripheral clusters.
In the case of BC, artefacts are observable forming a symmetrical pattern with a group of items regularly spaced between few dark dots.
These dark dots correspond to overlapping papers coupled with  the same reference and linked to no other reference.
Items at equi-distance from their neighbors are papers coupled with no other document and can be isolated using graph algorithms searching for weakly connected components.

The second major difference concerns cluster delineation.
For BC, the cluster density does not vary a lot with the location making difficult the identification of clusters' boundary.
For WBC, some boundaries are also difficult to identify.
Nevertheless, more clusters show a dense central area and a less dense peripheral area making easier cluster delineation.

These two differences are linked to the high-order neighborhood and the weighting of relevant references.
As a result, the number of isolated items is reduced by the larger number of references, increasing the overall connectivity.
In addition, the weighting impacts the formation of dense clusters by having more accurate coupling values leading to a better positioning.

\subsection{From Keywords to Clusters}

\subsubsection{Relevant Keywords Identification}

We compared in these paragraphs keywords ranked by \textit{raw} frequency and ranked by \textit{relative} frequency.
For each ranking, we list the $20$ first top keywords.
Using the \textit{raw} frequency ranking, we have:

\begin{displayquote}
	\textit{Payment, Computer science, Computer security, Computer network, Mathematical optimization, Economics, Internet privacy, Incentive, Microeconomics, Mathematics, Distributed computing, Marketing, The Internet, Database transaction, Mobile payment}
\end{displayquote}

and using the \textit{relative} frequency ratio $R(kw, V, V')$ obtained with Eq. \eqref{eq:occurrence_ratio}, keeping only keywords with at least $100$ occurrences, we have:

\begin{displayquote}
	\textit{Microeconomics, Incentive, Database transaction, Mobile payment, Payment service provider, Payment system, Mechanism design, Credit card, Actuarial science, Revenue, E-commerce, Smart card, Cash, Electronic money, Anonymity, Commerce, Common value auction, Cryptocurrency, Electronic cash, Blockchain}
\end{displayquote}

When looking at the former list, we can identify some words related to \textit{payment}, like \textit{Economics}, \textit{Microeconomics}, \textit{Marketing} and \textit{Mobile payment}.
All others are indirectly related to \textit{Payment}, as they correspond to concepts used for the development of different technologies, like \textit{Computer science} or \textit{Mathematics} which are very general concepts and reused in other domains.
In contrast, there are much more words related to \textit{Payment} in the second list, with only a few words where the direct connection is unclear like \textit{Incentive} or \textit{Actuarial science}.
Thus, the proposed ranking allows an uneven extraction of several pertinent keywords without efforts.

\subsubsection{Keywords Co-Localization}

The keywords in this second list will be studied by highlighting the papers' location where they occurs in the embedding.
The goal is to see if groups of papers sharing the same references also share the same keywords.

This measurement cannot be done using only the similarity values.
Two coupled papers are unlikely to share more keywords than non-coupled keywords because of the graph sparsity and the low number of keywords per document.

The keywords in the DBLP dataset are organized hierarchically.
A document is associated with general keywords and less general ones.
On average, all documents will share the general keywords with their neighbors.
When looking at the similarity between two documents' keywords, the contribution of infrequent keywords would be imperceptible.

As the number of keywords per document is limited  (on average $10$), not all valid keywords are associated with a document.
For instance, a \textit{Bitcoin} paper can include \textit{Cryptocurrency} or \textit{Blockchain} keywords.
However, its neighbors may be associated with one or none because they are almost equivalent and infrequent.
Therefore, the probability that coupled papers would share a specific keyword is very low for infrequent keywords.
This difference of predominance makes the direct evaluation of document similarity using keywords problematic.

Among direct neighbors, only a few would share the same specific keywords.
Over a larger group, we can expect to gather more papers associated with a particular keyword.
The use of an embedding allows considering indirect neighbors, increasing the probability to find neighbors with a particular keyword in common.

\begin{figure}
  \centering
  \includegraphics[width=\textwidth]{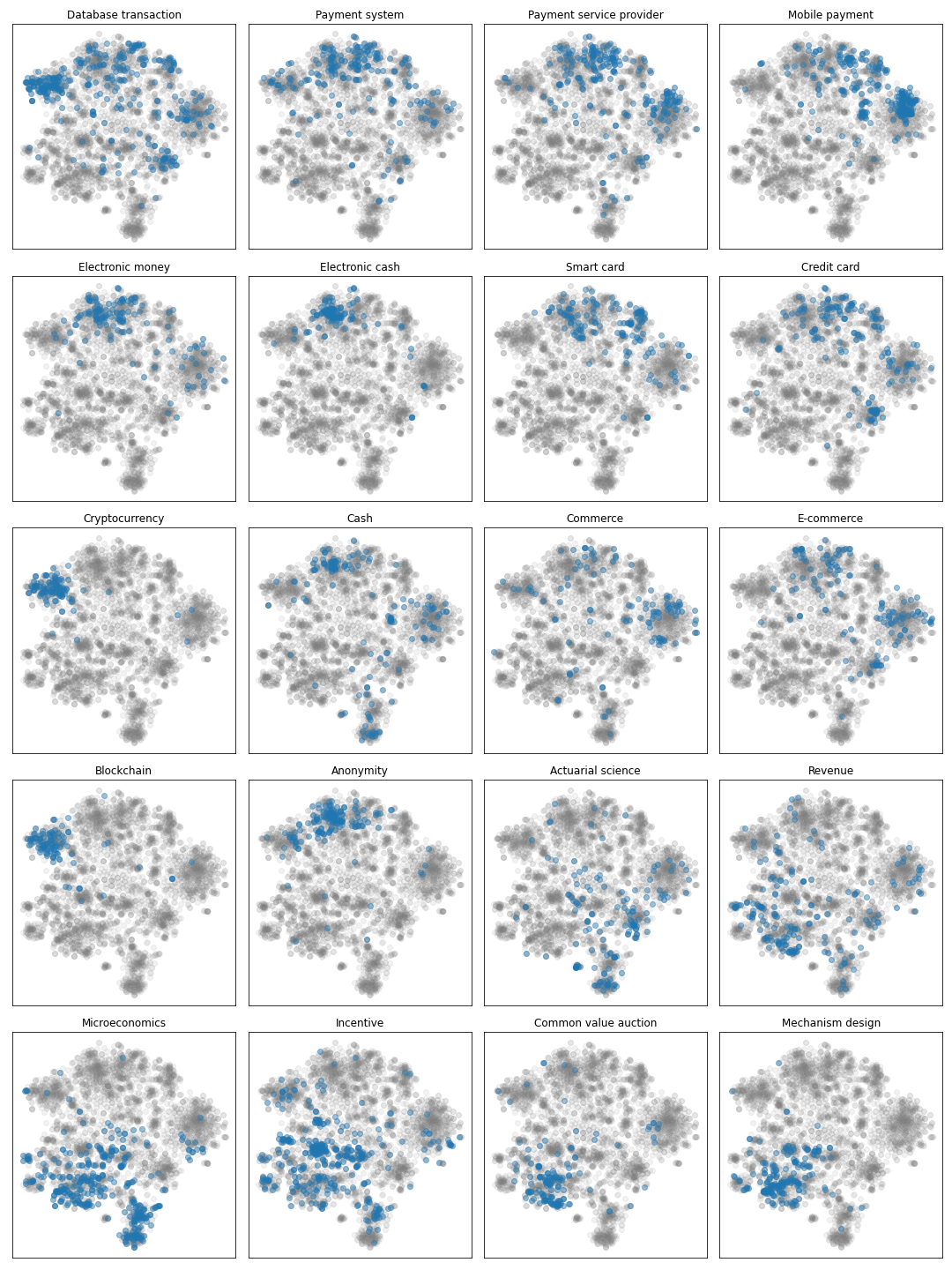}
  \caption{Local Map of Science for “Payment” papers with papers associated with a relevant keyword highlighted.
  Keywords are arranged based on their co-location similarity.
  }
  \label{fig:clusters}
\end{figure}

In  Fig. \ref{fig:clusters}, we show keywords location by highlighting papers associated with the current keyword.
As a general result, keywords tend to group into clustersis especially true for low-frequency keywords, like \textit{Blockchain} or \textit{Electronic cash}.
For keywords with larger frequency, like \textit{Database transaction}, \textit{Payment service provider}, or \textit{Microeconomics}, they occur in multiple clusters, possibly disjoint.
Finally, some keywords do not aggregate into dense clusters, such as \textit{Commerce}, \textit{E-commerce} or \textit{Revenue} which form diffused areas, partly due to their low frequency compared to other keywords.
Low-frequency keywords correspond to very specific topics, with a narrowed literature.
The ability to group and form a dense cluster is more likely for specific topics than for more general topics.
A broad topic can be sub-divided into subtopics.
A paper within a subtopic may have general references relative to the broad topic but others related to the subtopic.
Papers from different subtopics are likely to be weakly coupled, leading to the decomposition of a large topic into multiple clusters.

This map can also be used to study keywords relatedness by looking at their covered areas.
We can identify several groups of related keywords, like \textit{Blockchain} with \textit{Cryptocurrency};  \textit{Electronic money} with \textit{Electronic cash}, \textit{Cash} and \textit{Anonimity};   \textit{Common value auction} with \textit{Mechanism design}; \textit{Credit card} with \textit{Smart card}, etc.
Some keywords share exactly the same areas like \textit{Blockchain} and \textit{Cryptocurrency}, while others partially overlap, like \textit{Anonymity} overlapping with \textit{Blockchain} in the one side and \textit{Electronic cash} and related keywords for the other side.
The partial overlap shows that concepts are not hierarchically organized, as none dominate the other.
Nevertheless, a research area can be charaterized by a specific mixture of topics.

\subsection{From Clusters to Keywords}

In this experiment, we start from the papers' clusters to study keywords' groups instead of starting from keywords' clusters to identify papers' groups.

\subsubsection{Mean-Shift Partitioning}

\begin{figure}[h]
  \centering
  \includegraphics[width=0.65\textwidth]{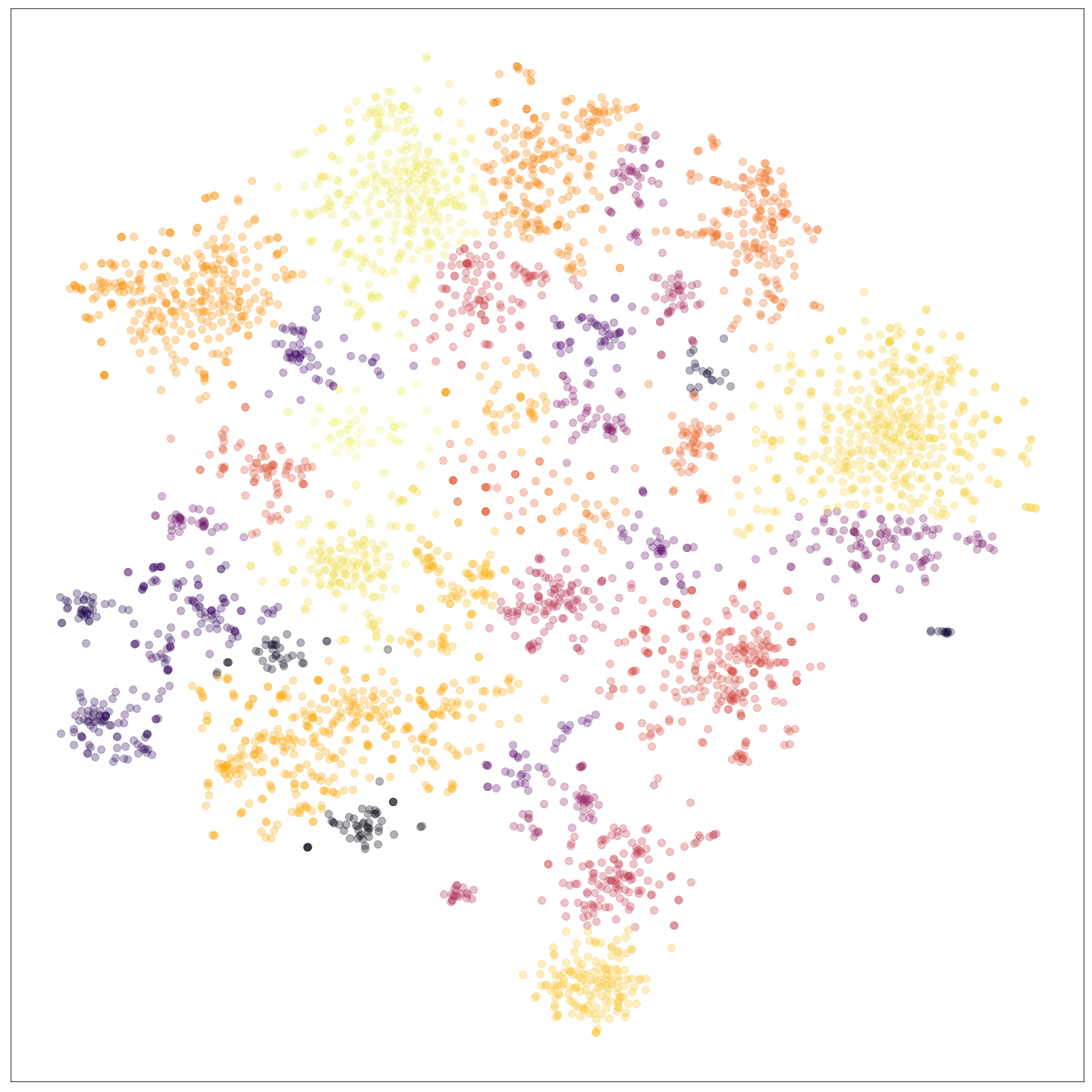}

  \caption{Mean-Shift clustering of the \textit{Payment} resulting embedding.}
\label{fig:cg_meanshift}
\end{figure}

The WBC embedding is clustered using the Mean-Shift algorithm, where $\sigma = 2.8$, corresponding to the average distance to the $30$ first nearest neighbors.
The partitioning results are presented in Fig. \ref{fig:cg_meanshift}, where each color is associated with a cluster.
The clustering led to the identification of 38 clusters, of which $34$ have at least $30$ documents.

The partitioning obtained led to the identification of clusters of various size and shape.
Compared to a $k$-Mean, the Mean-Shift algorithm has the advantage of not requiring as input the number of clusters and separating clusters by non-straight boundaries.
However, some potential "errors" exist, like the two clusters on the right (a yellow and a purple)  would be better merged.
In the middle of the embedding, different clusters are identified. As these documents correspond to weakly coupled items, their relevance and usability are unclear.

\subsubsection{Group Analysis}

Keywords are analyzed by ranking all keywords within a cluster using TF-IDF.
We assigned a keyword to a cluster based on its largest TF-IDF score to avoid multiple occurrences.
Some groups are presented in table \ref{tab:cg_keywords}, where keywords within a group are listed by decreasing score and keeping only those with several occurrences within the subset equal or greater than $20$. We kept the top-most relevant keywords as a matter of space.
Among the $34$ cluster candidates, we selected $28$ of them, the clusters containing the previously identified keywords and others with a sufficient number of relevant keywords for comparison.

Among the $20$ keywords in the second list, only \textit{Mechanism design} is not listed because of its poor TF-IDF score.
Only a minority of the keywords in the first list is listed, as their large number of occurrences leads to a very low IDF score.

The topic covered by each cluster is easy to identify.
Often, one of the keywords within the group is a good representative.
For example, for group $1$, \textit{Cryptocurrency} represents the main idea.
For group $11$, the \textit{Biometry} topic includes all the other sub-topics.
Group $16$ with \textit{Mathematical economics}.
For other groups, the idea can be summarized by using an external keyword.
For instance, group $3$ corresponds to payment technologies, including protocols and devices.
For group $20$, the idea can be summarized by user-related security.

There are some groups where the concept is unknown, such as  group $6$ or group $27$, which is partially due to the clustering and the single assignment of a keyword.
In addition, there are some keywords where the links to the group idea are unclear, such as \textit{Fuzzy logic} in group $25$, or \textit{Digital goods} in group $2$.

Words may have multiple meanings or cover different issues.
Within a group, the correct meaning becomes evidence.
For instance, \textit{Database transaction} refers to many things.
It could be the algorithms, ensuring ACID principles (Atomicity, consistency, isolation and durability), or refers to the recorded data.
Within the \textit{Cryptocurrency} cluster, the meaning becomes very clear, as one of the main blockchain issues is how to record the maximal number of transactions in a decentralized system without sacrificing security and correctness.
Associated with another cluster, the interpretation would have been different.
We assign each keyword to a cluster only once to avoid very long keywords lists; therefore we don't have another contextual example.
Nevertheless, for a more in-depth analysis, all relevant keywords must be considered.

This experiment allows identifying clusters' topics based on the main keywords.
Another experiment would be to study topics relationships by looking at the nearest clusters mixture.
Nonetheless, this approach is not recommended over a $t$-SNE embedding.
This embedding method  tries to preserve the local neighborhood, and distant neighbors are not organized specifically.
Therefore, conclusions concerning clusters' relationships might be inexact.

\begin{table}
  \caption{Table gathering keywords grouped by clusters and sorted by TF-IDF.}
  \label{tab:cg_keywords}

  \resizebox{\textwidth}{!}{
    \begin{tabular}{| c | c | c | c|}
      \hline
      Group 1                         & Group 2                  & Group 3                           & Group 4 \\
      \hline
      \textbf{Cryptocurrency}         & \textbf{Electronic cash} & \textbf{Mobile payment}           & Micropayment              \\
      \textbf{Blockchain}             & \textbf{Anonymity}       & \textbf{Payment service provider} & \textbf{Electronic money} \\
      Ledger                          & Blind signature          & Payment gateway                   & Cryptographic protocol    \\
      Smart contract                  & Trusted third party      & Mobile commerce                   & Hash chain                \\
      Communication channel           & Computer security        & \textbf{Credit card}              & \textbf{E-commerce}       \\
      Scalability                     & Money laundering         & Payment protocol                  & Hash function             \\
      Throughput                      & Digital goods            & Card security code                & 3-D Secure                \\
      Currency                        &                          & Digital signature                 & Payment order             \\
      \textbf{Database transaction}   &  & & \\
      \hline
      \hline

      Group  5                 & Group 6               & Group 7              & Group 8 \\
      \hline
      Near field communication & Multimedia            & Virtual machine      & Demand response         \\
      \textbf{Smart card}      & Digital currency      & Provisioning         & Electricity             \\
      Access control           & Welfare               & Cloud computing      & Smart grid              \\
      Contactless smart card   & \textbf{Cash}         &  Resource management & Control engineering     \\
      ATM card                 & Social media          &  Cost accounting     & Environmental economics \\
                               & Financial management  &  Resource allocation & Energy consumption      \\
    \hline
    \hline

    Group  9                 & Group 10               & Group 11        & Group 12 \\
    \hline
    Ubiquitous computing    & Network packet          & Biometrics      & Technical debt       \\
    Mobile device           & Wireless ad hoc network & Fingerprint     & Systems engineering  \\
    Mobile computing        & Wireless network        & Authentication  & Software             \\
    Embedded system         & Relay                   & Password        &  Process management  \\
    \textbf{Payment system} & Mobile telephony        & Usability       &  Risk analysis       \\
    World Wide Web          & Security token          &                 &  Management science  \\
    \textbf{Commerce}       & \textbf{Incentive}      &                 &  Empirical research  \\
    \hline
    \hline

    Group  13                 & Group 14                 & Group 15                      & Group 16 \\
    \hline
    Dividend                   & Operation management    & Online advertising            &  Mathematical economics       \\
    Econometrics               &  Supply chain           & \textbf{Common value auction} & Vickrey-Clarke-Groves auction \\
    Optimization problem       & \textbf{Microeconomics} & Advertising                   & Combinatorial auction         \\
    \textbf{Actuarial science} & Economics               & Bidding                       & Redistribution                \\
    Interest rate              & Information asymetry    & Valuation                     & Optimal mechanism             \\
    Transaction cost           & Profit                  & \textbf{Revenue}              & Social Welfare                \\
    Mathematics                & Industrial organization & Budget constraint             & \\

    \hline
    \hline
    Group  17                 & Group 18                 & Group 19                      & Group 20 \\
    \hline
    Net neutrality          & Cognitive radio             & Ransomware        & Android \\
    Bargaining problem      & Price of anarchy            & Malware           & Internet privacy \\
    Quality of service      & Cellular network            & Encryption        & Security analysis \\
    Internet access         & Stackelberg competition    & Monetization      & Information sensitivity \\
    Game theory             & Operator                    & Payment processor & Implementation  \\
    Profitability index     & Wireless                    &                   &  Vulnerability  \\
    Nash equilibrium        & Telecommunications network  &                   & Information security \\

    \hline
    \hline

    Group  21                 & Group 22                 & Group 23                      & Group 24 \\
    \hline
    Welfare economics  & Knowledge management   & Crowdsourcing   & Decision tree\\
    Linear programming & Medicine               & Data science    & Transaction data \\
    Schedule           & Information technology & Transparency    & Artificial neural networks \\
    Public transport   & Outsourcing            & Remuneration    & Machine learning \\
    Heuristics         & Decision support       &                 & Artificial intelligence \\
    Exploit            &                  &                 & Big data \\
    \hline
    \hline

    Group  25                 & Group 26                 & Group 27                      & Group 28 \\
    \hline
    Trade credit              & Technology acceptance model & Crowdsensing    & Peer-to-peer \\
    Economic order quantity   & Risk perception             & Rationality     & Upload         \\
    Inventory control         & Marketing                   & Reverse auction & Game theoretic \\
    Inventory shortage        & Psychology                  & Computation     & Broadcasting   \\
    Inflation                 & Perception                  &                 & Popularity \\
    Mathematical optimization & Developing country          &                 & Collusion  \\
    Vendor                    & & & Authorization \\
    Fuzzy logic               & & &  \\
    \hline

    \end{tabular}
  }
\end{table}

\section{Discussion}

\subsection{Would Bibliographic Coupling be Sufficient ?}

The main issue with BC is the lack of connectivity between papers, leading  mechanically to low similarity scores, and few coupling relationships.
Our experiment presented a subset of a paper belonging to the \textit{Payment} domain, a small world with many connections between sub-topics.
In the broader area, such as \textit{Machine Learning}, \textit{Computer security}, \textit{Information retrieval}, there can be more than  100,000 documents.
When randomly sampling papers from these domains, the probability of papers sharin references is low leading to many more artefacts than a narrowed field.

\subsection{Calibration of Random-Walk Path Length}

The WBC exploits weights computed using the random walk with restart assigned to $R^t(a)$ items.
The setup with $t=1$ corresponds to BC as only the first neighbors are explored with equal probability.
The value of $t$ is adjusted with the sparsity of the subset.
A very dense group of items requires a low value of $t$, around $2$ or $3$, while a very sparse structure needs a larger value of $4 \sim 5$
However, a larger $t$ impacts the computational cost; therefore using the lowest possible value is preferable.
We set $t=3$ for the experiments as it allowed us to test over sparse and less sparse subsets.

\subsection{Node Sampling}

We proposed a methodology for representing papers without prior knowledge.
The papers described here belong to the same subset.
Still, they are selected at random, without looking at age, number of received citations, number of references, language, conference or journal ranking.
The only bias was on the thematic, which narrows the scope and gives understandable relationships.
Other selection approaches can be proposed, such as looking at all papers published at a large conference or journal, for automatic field categorization.
Another would be to look at a university's papers, to study synergies between different research groups.

\subsection{Bottleneck Effect}

The proposed approach takes advantage of the random walk, where the influence of a node diffuse over the graph.
At the start, a weight of $1$ is dispatched between all first neighbors.
With a larger exploration distance, the amount dispatched is still $1$, or lower if some terminal nodes are reached.
Therefore, the system is almost conservative.
The amount gathered by a node depends on the distance.
For a low exploration distance, the weights gathered vanished because they are dispatched over many nodes.
After more exploration steps, the number of beneficiaries decreases because the set of papers in the past is smaller than in the recent years.
Therefore, old papers tend to concentrate the amount of diffused weights surpassing the amount collected by recent works.
The bottleneck effect can be avoided by slightly modifying the random walk by adding a decay factor $\alpha$, to consider the loss or transformation of information over time.

\subsection{Future versus Past}

As discussed earlier, an asymmetry exists between citation and referencing process.
Nonetheless, the proposed approach can  be used over citations rather than references, or a mix of the two.
The referencing approach focuses on “who share the same basis”, while with the citation approach, the focus would be “who are co-inspired”.
The conceptual difference might be relatively small, but more artefacts are likely to be present because of the power-law distribution of citations. However, this way might be a useful exploration path for papers impact evaluation.

\section{Conclusion}

In this paper, we proposed a methodology to extend bibliographic coupling to the more in-depth neighborhood.
The presented approach considers the indirect neighborhood of a specific paper, where their contribution is weighted using the Random Walk algorithm.
This approach allows to score neighborhood nodes based on their relative distance and accessibility from the parent paper, which corresponds to a form of relevance.
These weights are used to compute the similarity between two papers, using a weighted Cosine similarity.
The advantage against bibliographic coupling is the connectivity: much more similarity pairs are non-zero, which create a single connected component.
Following an embedding step with the $t$-SNE algorithm contains much fewer artefacts than bibliographic coupling, which leads to a better analysis of the sub-field relationships.
Because of the connectivity advantage, any papers could be represented on our map, without restriction on the number of citations.
This enlarges the possible candidate papers which can be used to build  local maps of science.

\bibliographystyle{unsrtnat}
\bibliography{references}  

\begin{thebibliography}{30}
\providecommand{\natexlab}[1]{#1}
\providecommand{\url}[1]{\texttt{#1}}
\expandafter\ifx\csname urlstyle\endcsname\relax
  \providecommand{\doi}[1]{doi: #1}\else
  \providecommand{\doi}{doi: \begingroup \urlstyle{rm}\Url}\fi

\bibitem[Garfield(2004)]{Garfield2004HistoriographicMO}
E.~Garfield.
\newblock Historiographic mapping of knowledge domains literature.
\newblock \emph{Journal of Information Science}, 30:\penalty0 119 -- 145, 2004.

\bibitem[Kessler(1963)]{Kessler1963BibliographicCB}
M.~M. Kessler.
\newblock Bibliographic coupling between scientific papers.
\newblock \emph{American Documentation}, 14:\penalty0 10--25, 1963.

\bibitem[Small(1973)]{Small1973CocitationIT}
H.~Small.
\newblock Co-citation in the scientific literature: A new measure of the
  relationship between two documents.
\newblock \emph{J. Am. Soc. Inf. Sci.}, 24:\penalty0 265--269, 1973.

\bibitem[Grauwin and Jensen(2011)]{Grauwin2011MappingSI}
S.~Grauwin and P.~Jensen.
\newblock Mapping scientific institutions.
\newblock \emph{Scientometrics}, 89:\penalty0 943--954, 2011.

\bibitem[Leydesdorff and Milojević(2013)]{leydesdorff2013scientometrics}
Loet Leydesdorff and Staša Milojević.
\newblock Scientometrics, 2013.

\bibitem[Ding(2011)]{Ding2011ScientificCA}
Ying Ding.
\newblock Scientific collaboration and endorsement: Network analysis of
  coauthorship and citation networks.
\newblock \emph{Journal of informetrics}, 5 1:\penalty0 187--203, 2011.

\bibitem[Boyack et~al.(2005)Boyack, Klavans, and Börner]{Boyack2005Backbone}
Kevin~W. Boyack, Richard Klavans, and Katy Börner.
\newblock Mapping the backbone of science.
\newblock \emph{Scientometrics}, 64\penalty0 (3):\penalty0 351--374, 2005.
\newblock URL
  \url{https://EconPapers.repec.org/RePEc:spr:scient:v:64:y:2005:i:3:d:10.1007_s11192-005-0255-6}.

\bibitem[Boyack and Klavans(2010)]{Boyack2010CocitationAB}
K.~Boyack and R.~Klavans.
\newblock Co-citation analysis, bibliographic coupling, and direct citation:
  Which citation approach represents the research front most accurately?
\newblock \emph{J. Assoc. Inf. Sci. Technol.}, 61:\penalty0 2389--2404, 2010.

\bibitem[Klavans and Boyack(2017)]{Klavans2017WhichTO}
R.~Klavans and K.~Boyack.
\newblock Which type of citation analysis generates the most accurate taxonomy
  of scientific and technical knowledge?
\newblock \emph{Journal of the Association for Information Science and
  Technology}, 68, 2017.

\bibitem[Price(1976)]{Price1976AGT}
D.~Price.
\newblock A general theory of bibliometric and other cumulative advantage
  processes.
\newblock \emph{J. Am. Soc. Inf. Sci.}, 27:\penalty0 292--306, 1976.

\bibitem[Redner(1998)]{Redner1998HowPI}
S.~Redner.
\newblock How popular is your paper? an empirical study of the citation
  distribution.
\newblock \emph{The European Physical Journal B - Condensed Matter and Complex
  Systems}, 4:\penalty0 131--134, 1998.

\bibitem[Merton(1968)]{Merton1968TheME}
R.~Merton.
\newblock The matthew effect in science.
\newblock \emph{Science}, 159:\penalty0 56 -- 63, 1968.

\bibitem[Allen et~al.(2013)Allen, Stanton, Pietro, and
  Moseley]{Allen2013SocialMR}
H.~Allen, T.~Stanton, F.~Di Pietro, and G.~Moseley.
\newblock Social media release increases dissemination of original articles in
  the clinical pain sciences.
\newblock \emph{PLoS ONE}, 8, 2013.

\bibitem[Gargouri et~al.(2010)Gargouri, Hajjem, Larivi{\`e}re, Gingras, Carr,
  Brody, and Harnad]{Gargouri2010SelfSelectedOM}
Y.~Gargouri, Chawki Hajjem, V.~Larivi{\`e}re, Y.~Gingras, L.~Carr, Tim Brody,
  and S.~Harnad.
\newblock Self-selected or mandated, open access increases citation impact for
  higher quality research.
\newblock \emph{PLoS ONE}, 5, 2010.

\bibitem[Abadi et~al.(2016)Abadi, Barham, Chen, Chen, Davis, Dean, Devin,
  Ghemawat, Irving, Isard, et~al.]{AbaBar16Tensorflow}
Mart{\'\i}n Abadi, Paul Barham, Jianmin Chen, Zhifeng Chen, Andy Davis, Jeffrey
  Dean, Matthieu Devin, Sanjay Ghemawat, Geoffrey Irving, Michael Isard, et~al.
\newblock Tensorflow: A system for large-scale machine learning.
\newblock In \emph{OSDI}, volume~16, pages 265--283, 2016.

\bibitem[Tahamtan and Bornmann(2019)]{Tahamtan2019WhatDC}
I.~Tahamtan and L.~Bornmann.
\newblock What do citation counts measure? an updated review of studies on
  citations in scientific documents published between 2006 and 2018.
\newblock \emph{Scientometrics}, 121:\penalty0 1635 -- 1684, 2019.

\bibitem[Kerckhoffs(1883)]{lacryptomilitaire}
A.~Kerckhoffs.
\newblock La cryptographie militaire.
\newblock \emph{Journal des sciences militaires}, IX:\penalty0 161--191, 1883.

\bibitem[Jolliffe(2005)]{PCA}
Ian Jolliffe.
\newblock \emph{Principal Component Analysis}.
\newblock John Wiley \& Sons, Ltd, 2005.
\newblock ISBN 9780470013199.
\newblock \doi{10.1002/0470013192.bsa501}.
\newblock URL \url{http://dx.doi.org/10.1002/0470013192.bsa501}.

\bibitem[Garfield and Zuckerman(1980)]{Garfield1980IndependentDiscovery}
E.~Garfield and H.~Zuckerman.
\newblock Multiple independent discovery and creativity in science.
\newblock \emph{Essays of an Information Scientist}, 4:\penalty0 1979--1980,
  1980.

\bibitem[Jarneving(2007)]{Jarneving2007BibliographicCA}
Bo~Jarneving.
\newblock Bibliographic coupling and its application to research-front and
  other core documents.
\newblock \emph{J. Informetrics}, 1:\penalty0 287--307, 2007.

\bibitem[Leydesdorff and R{\`a}fols(2009)]{Leydesdorff2009AGM}
L.~Leydesdorff and I.~R{\`a}fols.
\newblock A global map of science based on the isi subject categories.
\newblock \emph{J. Assoc. Inf. Sci. Technol.}, 60:\penalty0 348--362, 2009.

\bibitem[Fujita et~al.(2012)Fujita, Kajikawa, Mori, and
  Sakata]{Fujita2012DetectingRF}
K.~Fujita, Y.~Kajikawa, Junichiro Mori, and I.~Sakata.
\newblock Detecting research fronts using different types of weighted citation
  networks.
\newblock \emph{2012 Proceedings of PICMET '12: Technology Management for
  Emerging Technologies}, pages 267--275, 2012.

\bibitem[Huang and Chang(2013)]{Huang2013DetectingRF}
Mu-Hsuan Huang and Chia-Pin Chang.
\newblock Detecting research fronts in oled field using bibliographic coupling
  with sliding window.
\newblock \emph{Scientometrics}, 98:\penalty0 1721--1744, 2013.

\bibitem[Colavizza et~al.(2018)Colavizza, Boyack, van Eck, and
  Waltman]{Colavizza2018TheCT}
Giovanni Colavizza, K.~Boyack, Nees~Jan van Eck, and L.~Waltman.
\newblock The closer the better: Similarity of publication pairs at different
  cocitation levels.
\newblock \emph{Journal of the Association for Information Science and
  Technology}, 69, 2018.

\bibitem[Huang et~al.(2004)Huang, yun Chiang, and
  Chen]{Huang2004ConstructingAP}
Mu-Hsuan Huang, Li~yun Chiang, and Dar-Zen Chen.
\newblock Constructing a patent citation map using bibliographic coupling: A
  study of taiwan's high-tech companies.
\newblock \emph{Scientometrics}, 58:\penalty0 489--506, 2004.

\bibitem[Yan and Ding(2012)]{Yan2012ScholarlyNS}
Erjia Yan and Ying Ding.
\newblock Scholarly network similarities: How bibliographic coupling networks,
  citation networks, cocitation networks, topical networks, coauthorship
  networks, and coword networks relate to each other.
\newblock \emph{J. Assoc. Inf. Sci. Technol.}, 63:\penalty0 1313--1326, 2012.

\bibitem[Page et~al.(1999)Page, Brin, Motwani, and Winograd]{Page1999ThePC}
Lawrence Page, S.~Brin, R.~Motwani, and T.~Winograd.
\newblock The pagerank citation ranking : Bringing order to the web.
\newblock In \emph{WWW 1999}, 1999.

\bibitem[Tang et~al.(2008)Tang, Zhang, Yao, Li, Zhang, and Su]{DBLP}
Jie Tang, Jing Zhang, Limin Yao, Juanzi Li, Li~Zhang, and Zhong Su.
\newblock Arnetminer: Extraction and mining of academic social networks.
\newblock In \emph{KDD'08}, pages 990--998, 2008.

\bibitem[van~der Maaten and Hinton(2008)]{vanDerMaaten2008}
Laurens van~der Maaten and Geoffrey Hinton.
\newblock Visualizing data using {t-SNE}.
\newblock \emph{Journal of Machine Learning Research}, 9:\penalty0 2579--2605,
  2008.
\newblock URL \url{http://www.jmlr.org/papers/v9/vandermaaten08a.html}.

\bibitem[Cheng(1995)]{Mean_shift}
Yizong Cheng.
\newblock Mean shift, mode seeking, and clustering.
\newblock \emph{IEEE Trans. Pattern Anal. Mach. Intell.}, 17\penalty0
  (8):\penalty0 790–799, August 1995.
\newblock ISSN 0162-8828.
\newblock \doi{10.1109/34.400568}.
\newblock URL \url{https://doi.org/10.1109/34.400568}.

\end{thebibliography}






\end{document}